\documentstyle[prd,aps,epsf,12pt]{revtex}
\begin{document}

\title{
\hfill{\small { TRI-PP-00-23}}\\
\hfill{\small {MKPH-T-00-03}}\\
\hfill{\small { FZJ-IKP(TH)-2000-03}}\\[0.2cm]
Radiative pion capture by a nucleon}

\author{Harold W. Fearing$^a$\footnote{email: fearing@triumf.ca},
	Thomas R. Hemmert$^b$\footnote{email: th.hemmert@fz-juelich.de},
        Randy Lewis$^c$\footnote{email: randy.lewis@uregina.ca}
    and Christine Unkmeir$^d$\footnote{email: unkmeir@kph.uni-mainz.de}}

\address{$^a$ TRIUMF, 4004 Wesbrook Mall, Vancouver,
	      BC, Canada V6T 2A3}
\address{$^b$ Forschungszentrum J\"ulich, Institut f\"ur Kernphysik (Th),
	      D-52425 J\"ulich, Germany }
\address{$^c$ Department of Physics, University of Regina, Regina, SK, Canada
	      S4S 0A2}
\address{$^d$ Institut f\"ur Kernphysik, Johannes Gutenberg-Universit\"at,
              D-55099 Mainz, Germany}

\date{May 21, 2000}

\maketitle

\thispagestyle{empty}

\vspace{1cm}

\begin{abstract}
The differential cross sections for $\pi^- p \rightarrow \gamma n$ and $\pi^+ n
\rightarrow \gamma p$ are computed up to $O(p^3)$ in heavy baryon chiral
perturbation theory (HBChPT).  The expressions at $O(p)$ and $O(p^2)$ have no
free parameters.  There are three unknown parameters at $O(p^3)$, low energy
constants of the HBChPT Lagrangian, which are determined by fitting to
experimental data.  Two acceptable fits are obtained, which can be separated by
comparing with earlier dispersion relation calculations of the inverse
process. Expressions for the multipoles, with emphasis on the p-wave
multipoles, are obtained and evaluated at threshold. Generally the results
obtained from the best of the two fits are in good agreement with the
dispersion relation predictions.
\end{abstract}
\newpage

\section{Introduction}

Radiative pion capture by a nucleon is one of the obvious reactions to use as a
testbed for heavy baryon chiral perturbation theory (HBChPT).  For charged
pions, the reaction begins at $O(p)$, which is leading order in HBChPT, and it
is known that the $O(p^3)$ result for s-wave multipole is in reasonable
agreement with most measurements \cite{BKM}. The p-wave multipoles however seem
never to have been calculated. This is in contrast to the neutral pion case
where both s- and p-wave multipoles have been extensively discussed 
\cite{pizero}. A calculation beyond the s-wave provides insight into the
convergence of the chiral expansion and also serves to determine some of the
HBChPT parameters that are required for other reactions, such as radiative muon
capture by a nucleon, where the existing experimental data are in surprising
disagreement with theoretical expectations \cite{rmc}. Thus an investigation of
the p-wave multipoles in the charged case is a useful thing to do and is the
primary aim of this work.

In the present work, the only explicit fields in the chiral Lagrangian are the
pions and nucleons.  Other physical particles will enter the calculation
through their implicit contributions to the Lagrangian's parameters (LEC's).
For some reactions it is advantageous to include the $\Delta(1232)$ explicitly,
as done for example in Ref. \cite{Thomas}, and it is possible that this could
be a useful approach for radiative pion capture as well, once one goes away
from threshold.  However, it is consistent to absorb such resonances in the
LEC's and we shall see that for the present reaction a reasonable fit to the
data can be obtained when the $\Delta(1232)$ is left implicit in the HBChPT
parameters.

Experimental data for the $\pi^- p \rightarrow \gamma n$ differential cross
section was reported fifteen years ago from a TRIUMF experiment at beam
energies of $T_\pi = 27.4$ and 39.3 MeV \cite{Salomon}. A recent TRIUMF
experiment has taken data at $T_\pi = 9.88$, 14.62 and 19.85 MeV
\cite{newdata}.  There is also very recent data \cite{Korkmaz} for the inverse
reaction $\gamma p \rightarrow n \pi^+$ taken very near threshold at $T_\gamma
\simeq$ 153 MeV corresponding to $T_\pi \simeq 3$ MeV. In this study, we will
not attempt to apply HBChPT to energies above 40 MeV.

There are at least two modern theoretical discussions of radiative charged-pion
capture (both discussions actually address the inverse reaction: charged-pion
photoproduction).  One is an HBChPT study of the s-wave at threshold by
Bernard, Kaiser and Mei{\ss}ner\cite{BKM}, and another is a dispersion
theoretical analysis of s- and p-waves by Hanstein, Drechsel and
Tiator\cite{Tiator}.  The present work goes beyond threshold and also
explicitly computes the p-wave multipoles.  The comparison of our work to the
threshold results of Ref. \cite{Tiator} is found to be quite interesting and to
provide a useful constraint on our results.

In section \ref{sec2}, we establish the general expressions for kinematics,
multipoles and the differential cross section. Section \ref{sec3} discusses the
HBChPT calculation and section \ref{sec4} presents and discusses our results,
both at threshold and in general.  Section \ref{sec5} contains a summary of
what has been learned from this effort, and what the next steps could be.

\section{Kinematics and Multipoles}\label{sec2}

In radiative charged-pion capture by a nucleon, a low energy $\pi^\pm$ with
four-momentum $q^\mu =(E_\pi ,\vec{q})$ in the center-of-mass system gets
absorbed by a slowly moving nucleon of mass $m_N$. In the final state, one
observes a recoiling nucleon and a low energy photon with polarization
four-vector $\epsilon^\mu=(\epsilon_0,\vec{\epsilon})$ and four-momentum $k^\mu
=(\omega ,\vec{k})$. The pion's center-of-mass energy is related to $s$, the
square of the total energy in the center of mass, and to $T_\pi$, the kinetic
energy in the lab frame by
\begin{equation}
E_\pi = \frac{s+m_\pi^2-m_N^2}{2\sqrt{s}} = 
\frac{m_\pi^2+m_N(m_\pi+T_\pi)}{\sqrt{(m_N+m_\pi)^2+2m_N{T}_\pi}}
   \; ,
\end{equation}
where $m_\pi$ and $m_N$ are respectively the pion and nucleon masses.
The analogous formulas for the photon energy in the center of mass are 
\begin{equation}
\omega =\frac{s-m_N^2}{2\sqrt{s}} = 
\frac{m_N T_\gamma}{\sqrt{m_N^2+2m_N{T}_\gamma}}
   \; ,
\end{equation}
where $T_\gamma$ is the corresponding laboratory gamma energy for the inverse
process. All energy dependence will be expressed via the pion energy in the
center-of-mass system. For the energy of the final state photon we therefore
employ
\begin{equation}
\omega=E_\pi-\frac{m_{\pi}^2}{2m_N} + \frac{E_\pi m_\pi^2}{2 m_{N}^2} +
{\cal O}(1/m_{N}^3) \; .
\end{equation}

The differential cross section for
the pion capture process in the center-of-mass frame is,
\begin{equation}\label{diffxsecpi}
   \frac{{\rm d}\sigma^{\pi N \rightarrow \gamma N}}{{\rm d}\Omega_\gamma} = 
   \frac{\omega}{|\vec{q}|}
   \frac{1}{2}\sum_{\rm pol's} 
   \left| {\cal M} \right|^2 \; ,
\end{equation}
and that for the inverse (photoproduction) reaction is
\begin{equation}\label{diffxsecga}
   \frac{{\rm d}\sigma^{\gamma N \rightarrow \pi N}}{{\rm d}\Omega_\pi} = 
   \frac{|\vec{q}|}{\omega}
   \frac{1}{4}\sum_{\rm pol's} 
   \left| {\cal M} \right|^2 \; ,
\end{equation}
where ${\cal M}$ is the amplitude defined below.  Notice that
Eqs. (\ref{diffxsecpi}) and (\ref{diffxsecga}) explicitly contain the average
over initial and sum over final spins and polarizations and that the two cross
sections are related by the usual detailed balance relation.

Essentially all previous work has dealt with the inverse, photoproduction,
process, $\gamma N \rightarrow \pi N$ and the conventions for that process are
by now well established. Thus in the Coulomb gauge with $\epsilon_0=0$ and the
transversality condition $\vec{\epsilon} \cdot \vec{k} = 0$ the amplitude for
that process can be written in terms of the T-matrix as \cite{chew,berends}
\begin{eqnarray}\label{gamamp}
{\cal M}^{\gamma N \rightarrow  \pi N} =
\frac{m_N}{4\pi \sqrt{s}} \; T \cdot \epsilon &=&
F_1(E_\pi, x) \; i \chi^\dagger \vec{\sigma} \cdot \vec{\epsilon} \; \chi +
F_2(E_\pi, x) \; \chi^\dagger \vec{\sigma} \cdot \hat{q} \; \vec{\sigma}
\cdot \left( \hat{k} \times \vec{\epsilon} \right) \chi + \nonumber \\
& & F_3(E_\pi, x) \; i \chi^\dagger \vec{\sigma} \cdot \hat{k} \; \vec{
\epsilon} \cdot \hat{q} \; \chi +
F_4(E_\pi, x) \; i \chi^\dagger \vec{\sigma} \cdot \hat{q} \; \vec{\epsilon}
\cdot \hat{q} \; \chi \; , \label{eq:T}
\end{eqnarray}
where 
$\sigma^i$ is a Pauli matrix in spin space between the two-component spinors of
the incoming/outgoing nucleon ($\chi /\chi^\dagger$), $\epsilon$ is the photon
polarization vector and $x=\cos\theta$ corresponds to the cosine of the angle
between the photon and the pion momenta.

Furthermore, each structure amplitude $F_i(E_\pi, x)$ ($i$=1,2,3,4) can be
decomposed into three isospin channels ($a$=1,2,3)
\begin{equation}
F_i^a(E_\pi, x)=F_i^{\it (-)}(E_\pi, x) \; i \epsilon^{a3b} \tau^b
+F_i^{\it (0)}(E_\pi, x) \; \tau^a + F_i^{\it (+)}(E_\pi, x) \; \delta^{a3}
\; ,
\end{equation}
where $\tau^a$ denotes a Pauli matrix in isospin space. The physical structure
amplitudes are then obtained from the linear combinations
\begin{eqnarray}
F_i^{\gamma n \rightarrow \pi^-p } &=&
   \sqrt{2}\left[F_i^{\it (0)}-F_i^{\it (-)}\right] \; , \\
F_i^{\gamma p \rightarrow \pi^+n } &=&
   \sqrt{2}\left[F_i^{\it (0)}+F_i^{\it (-)}\right] \; .
\end{eqnarray}

The full physics content of this process is encoded in the four structure
amplitudes $F_i$, which are complicated functions of $E_\pi$ and $\theta$, and
in the amplitude of Eq. (\ref{eq:T}), the square of which is used to get the
cross section. However it may be more intuitive to discuss the underlying
physics in terms of a multipole decomposition.  The HBChPT formalism which we
are employing in the following sections involves an expansion in terms of the
pion energy divided by a scale of approximately 1 GeV, i.e. it is only reliable
in a kinematic region of low energy pions.  With this in mind we
restrict the multipoles we consider to s- and p-waves only.  They can be found
from the $F$-amplitudes via \cite{chew,berends}
\begin{eqnarray}\label{defE0pl}
E_{0+}\left(E_\pi\right)&=&\int_{-1}^1dx \left\{
	 \frac{1}{2}F_1\left(E_\pi,x\right)-\frac{1}{2} x
	 F_2\left(E_\pi,x\right)+\frac{1}{6}\left[1-
	 P_2(x)\right]F_4\left(E_\pi,x\right)\right\}, \\
M_{1+}\left(E_\pi\right)&=&\int_{-1}^1dx \left\{ \frac{1}{4}x
	 F_1\left(E_\pi,x
	 \right)-\frac{1}{4} P_2(x) F_2\left(E_\pi,x\right)+
	 \frac{1}{12}\left[P_2(x)-1\right]F_3\left(E_\pi,x\right)
	 \right\}, \\
M_{1-}\left(E_\pi\right)&=&\int_{-1}^1dx \left\{ -\frac{1}{2} x
	 F_1\left(E_\pi,x
	 \right)+\frac{1}{2}F_2\left(E_\pi,x\right)+\frac{1}{6}
	 \left[1-P_2(x)\right]F_3\left(E_\pi,x\right)\right\}, \\
E_{1+}\left(E_\pi\right)&=&\int_{-1}^1dx \left\{ \frac{1}{4} x
	 F_1\left(E_\pi,x \right)-\frac{1}{4} P_2(x)
	 F_2\left(E_\pi,x\right)+\frac{1}{12}\left[1-P_2(x)\right]
	 F_3\left(E_\pi,x\right) \right.\nonumber \\
      && \left.+\frac{1}{10}\left[x-P_3(x)\right]
	 F_4\left(E_\pi,x\right)\right\}, \label{defE1pl}
\end{eqnarray}
with the $P_i(x),\;i\geq 2$ being Legendre polynomials.

The formulas above are those conventionally defined for the photoproduction
reaction $\gamma N \rightarrow \pi N$, whereas we are interested particularly
in the capture process $\pi N \rightarrow \gamma N$. The cross sections for
these two processes are related trivially by the detailed balance equation
arising from Eqs.(\ref{diffxsecpi}) and (\ref{diffxsecga}). The relation
between the amplitudes is however more complicated, arising from time reversal
and depending explicitly on the phases of the parts of the amplitude. In our
conventions we find (up to a possible overall, and thus irrelevant phase)
\begin{equation}\label{trev}
{\cal M}^{\pi N \rightarrow \gamma N} = 
-[{\cal M}^{\gamma N \rightarrow \pi N}]^*.  
\end{equation}

If we apply Eq. (\ref{trev}) to Eq. (\ref{gamamp}) to get the amplitude for
pion capture the structure functions $F_i$ attract various phases and a complex
conjugate and the order of the structures corresponding to $F_2$ is reversed.
Putting the $F_2$ structures back in the original order generates extra terms
and makes some of the coefficients of the four independent structures linear
combinations of the $F_i$. Thus if we were to define the amplitude for the pion
capture reaction to be of the original general form of Eq. (\ref{gamamp}) then
the $F_i$ for pion capture will be linear combinations, complex conjugated,
with various phase changes, of the $F_i$ for photoproduction. An alternative,
and probably more sensible choice, is to define the amplitude for the capture
reaction via the action of Eq. (\ref{trev}) on the definition used for the
photoproduction direction. This eliminates the problem of linear combinations,
but still leaves the two sets of $F_i$ related by a complex conjugate and
various phase changes.

The third alternative, which is the one we adopt, is to just do the calculation
for the photoproduction direction in the first place, and then make the
connection to the pion capture direction at the level of the cross section.
This has the advantage of keeping a close connection with the conventions and
the large body of previous work dealing with photoproduction.  Thus the
formulas for the $F_i$ which we quote, and more importantly those for the
multipoles, are actually for the $\gamma N \rightarrow \pi N$ direction. This
means for example that our numerical results for the multipoles can be compared
directly and without ambiguity with the dispersion relation calculation for
photoproduction of Ref. \cite{Tiator}, even though the parameters are being
fixed primarily by the pion capture data.

\section{The HBChPT Calculation}\label{sec3}

The HBChPT Lagrangian is ordered in powers of momenta and pion
masses, which are small compared to both the chiral scale, $4{\pi}F$,
and the nucleon mass, $m_N$,
\begin{equation}
   {\cal L}_{\pi{N}} = {\cal L}^{(1)}_{\pi{N}} + {\cal L}^{(2)}_{\pi{N}}
		     + {\cal L}^{(3)}_{\pi{N}} + \ldots \; .
\end{equation}
The lowest-order Lagrangian is
\begin{equation}
   {\cal L}^{(1)}_{\pi{N}} = \bar{N}_v(iv{\cdot}\nabla+g_AS{\cdot}u)N_v
\end{equation}
where \cite{EM,FLMS}
\begin{eqnarray}
 N_v(x) &=& \exp\left[im_{0N}v{\cdot}x\right]\frac{1}{2}(1+v\!\!\!/)\psi(x),\\
   S_\mu &=& \frac{i}{2}\gamma_5\sigma_{\mu\nu}v^\nu, \\
   u_\mu &=& iu^\dagger(\partial_\mu-ir_\mu)u
	    -iu(\partial_\mu-i\ell_\mu)u^\dagger, \\
   \nabla_\mu &=& \partial_\mu + \Gamma_\mu - iv_\mu^{(s)}, \\
   \Gamma_\mu &=& \frac{1}{2}\left[u^\dagger(\partial_\mu-ir_\mu)u
		  +u(\partial_\mu-i\ell_\mu)u^\dagger\right],
\end{eqnarray}
with $m_{0N}$ and $g_A$ being the lowest-order nucleon mass and axial
coupling respectively.	The external photon field is included via
$r_\mu = \ell_\mu = -(e/2)\tau^3A_\mu$,
and $u$ is a nonlinear representation of the pion fields, for example
\begin{equation}
   u = \exp\left[\frac{i}{2F_0}\left(\begin{tabular}{cc}
		 $\pi^0$ & $\sqrt{2}\pi^+$ \\
		 $\sqrt{2}\pi^-$ & $-\pi^0$ \end{tabular}\right)\right].
\end{equation}
The parameter $F_0$ corresponds to the pion decay constant in the chiral limit
(normalized so that the physical value $F=92.4$ MeV).

The higher-order Lagrangians ${\cal L}^{(n)}_{\pi{N}}$
will be written in the notation of Ecker and Moj\v{z}i\v{s} \cite{EM} and are
exactly the same as those used in Ref. \cite{FLMS}.
Results for the multipoles in the present work depend on four combinations
of parameters
from ${\cal L}^{(3)}_{\pi{N}}$, namely $b_{10}$, $b_{19}$, $b_{21}^r(\mu)$
and $2b_{22}^r(\mu) + b_{23}$, where $\mu$ is the renormalization scale.
The numerical values of $b_{19}$ and $b_{23}$ were determined in
Ref. \cite{FLMS}.
The three remaining parameters, $b_{10}$, $b_{21}^r(\mu)$ and $b_{22}^r(\mu)$,
will be determined in the present work.

The calculation requires an evaluation of tree-level and one-pion-loop
diagrams, which can be organized into four classes depending on whether the
radiated photon is emitted from the initial nucleon, the final nucleon, the
pion, or from the $\pi{N}N$ vertex.  The calculation was performed in a general
gauge (and is fully gauge invariant). While this meant more work, the ability
to check gauge invariance provided a very important tool for eliminating errors
in what was an algebraically complex calculation. The result was then reduced
to the special case of $v{\cdot}\epsilon = 0$.  In this gauge, only one of the
four classes of diagrams has any dependence on the unknown HBChPT parameters,
$b_{10}$, $b_{21}^r(\mu)$ and $b_{22}^r(\mu)$, namely photon emission from the
$\pi{N}N$ vertex.

Adding all contributions together gives the amplitude of Eq. (\ref{eq:T}) with
the structure amplitudes, $F_i(E_\pi,x)$, given explicitly in Appendix
\ref{appA}.  Although only charged-pion processes are discussed in this work,
the calculation was actually performed for general isospin.  We have verified
that the $\pi^0$ amplitudes agree with Ref.~\cite{pizero}.

\section{Results}\label{sec4}
\subsection{The differential cross section}

Using our calculation from the previous section with Eq.~(\ref{diffxsecpi}) or
Eq.~(\ref{diffxsecga}) and the ${\cal M}$ of Eq.~(\ref{gamamp}) and the F's of
the Appendix, we can immediately compute the differential cross section.  At
$O(p)$ and $O(p^2)$ the result is completely determined, whereas at $O(p^3)$ it
depends on three unknown parameters, which will now be determined via a
least-squares fit to the experimental data.

Ref. \cite{newdata} provides 11 measurements of the differential cross section
for $\pi^- p \rightarrow \gamma n$ at $T_\pi = 9.88$, 14.62 and 19.85 MeV and
Ref. \cite{Salomon} provides an additional 16 measurements at $T_\pi = 27.4$
and 39.3 MeV. A further 8 measurements, these for the inverse reaction $\gamma
p \rightarrow n \pi^+$ very near threshold ($T_\pi \simeq 3$ MeV), come from
Ref. \cite{Korkmaz}.  We have performed fits to several subsets of this set of
data, as well as to the complete set.  A comparison of these fits allows us to
check for consistency among the data sets and also for a possible breakdown of
the HBChPT form as $T_\pi$ increases.

The values of the three fitted parameters are given in Table \ref{tab:bs}.  It
is reassuring to see that within the uncertainties all of the various data sets
lead to the same numerical values for these parameters, though the fit becomes
more stable and the uncertainties smaller as we increase the number of data
points included in the fit.  It should also be noted that each of our
least-squares fits actually finds two sets of parameters, characterized by
nearly identical values of $b_{21}^r$ and $b_{22}^r$ but quite different values
of $b_{10}$, depending on where the least-squares routine begins in
parameter space.  This presumedly reflects the fact that the cross section is
quadratic in the $b_i$'s and that the data is not sufficiently good to
distinguish the two solutions. We refer to these two minima in parameter-space
as ``A'' and ``B'', and then label our solution sets as A($n$) and B($n$),
where $n$ is the number of experimental measurements used in the fit.  For the
various subsets of pion capture data, A($n$) and B($n$) give essentially
indistinguishable $\chi^2$ values and differential cross sections. Addition of
the very low energy photoproduction data of Ref. \cite{Korkmaz} produces a
small improvement in the $\chi^2$ of the A(35) solution relative to the B(35)
one.  The two solutions can be distinguished, however, by their quite different
values of $b_{10}$ and also by the different individual p-wave multipoles, as
will be discussed below.

The results of our best fits to the cross section data are shown in
Fig. \ref{plots}, along with the parameter-free $O(p)$ and $O(p^2)$
calculations and the experimental data.  As these plots indicate, the $O(p)$
calculation disagrees with the data.  $O(p^2)$ contributions reduce the
discrepancy, but do not eliminate it.  The $O(p^3)$ terms are necessary for a
good fit to the data. The $O(p)$ terms clearly dominate (note the suppressed
zero in the plots), but the contributions of $O(p^2)$ and $O(p^3)$ are
comparable at most angles. For $\gamma n \leftrightarrow \pi^- p$ the two
contributions seem to add, whereas for $\gamma p \leftrightarrow \pi^+ n$ they
have opposite signs and tend to cancel. The fact that the $O(p^2)$ and $O(p^3)$
terms are more or less equal may raise some concern that the HBChPT expansion
has not yet fully converged at $O(p^3)$.  This point can also be made from
Table \ref{tab:bs}, which gives the values of the three parameters that were
determined in the fits.  For a nicely converging chiral expansion that just
contains pions and nucleons as effective degrees of freedom, one probably would
have expected each of the $b_i$ to acquire values near unity.  The fact that we
find values somewhat larger than this perhaps can be seen as an indication of
the role of explicit matter fields like the $\Delta$ isobar and vector
mesons. The discussion of such issues, however, has to be delayed to a future
communication. Here we only provide the first step and fix the contact terms
numerically at the scale $\mu=m_N$. Note however that the value of $b_{10}$
obtained from the A (but not the B) solution is quite consistent in magnitude
with the value of the parameter $b_P$, which is a linear combination of
$b_{10}$ and $b_{9}$, obtained in Ref. \cite{pizero} by fitting $\pi^0$ data.

\subsection{Threshold Results}

Expressions are simplified somewhat at threshold, that is in the limit in which
the pion kinetic energy $T_\pi$ goes to zero.  Using $M_{1+} =
\omega|\vec{q}|m_{1+}$, $M_{1-} = \omega|\vec{q}|m_{1-}$, and $E_{1+} =
\omega|\vec{q}|e_{1+}$, the multipoles, given for the photoproduction process
$\gamma + N \rightarrow \pi + N$, follow directly from
Eqs. (\ref{defE0pl}-\ref{defE1pl}) and the expressions for the $F$'s given in
the Appendix.  The resulting expressions are given purely in terms of physical
quantities. For the ${\it (0)}$ isospin channel we obtain
\begin{eqnarray}\label{thresh1}
E_{0+}^{\it (0)}(m_\pi) &=& \frac{m_N}{4\pi(m_N+m_\pi)}\frac{eG_A}{2F}
       \left[-\frac{m_\pi}{2m_N}+\frac{m_\pi^2}{4m_N^2}(\mu_p+\mu_n)\right]
       \; , \\
m_{1+}^{\it (0)}(m_\pi) &=& \frac{m_N}{4\pi(m_N+m_\pi)}\frac{eG_A}{2F}
       \left[\frac{(\mu_p+\mu_n)}{6m_\pi{m}_N}-\frac{1}{12m_N^2}
       +\frac{(\mu_p+\mu_n)}{6m_N^2}
       +\frac{2b_{10}}{3G_A(4\pi{F})^2}\right] \; , \\
m_{1-}^{\it (0)}(m_\pi) &=& \frac{m_N}{4\pi(m_N+m_\pi)}\frac{eG_A}{2F}
       \left[-\frac{(\mu_p+\mu_n)}{3m_\pi{m}_N}+\frac{7}{24m_N^2}
       -\frac{(\mu_p+\mu_n)}{3m_N^2}
       +\frac{2b_{10}}{3G_A(4\pi{F})^2}\right] \; , \\
e_{1+}^{\it (0)}(m_\pi) &=& \frac{m_N}{4\pi(m_N+m_\pi)}\frac{eG_A}{2F}
       \frac{1}{24m_N^2} \; ,
\end{eqnarray}
and for the ${\it (-)}$ isospin channel 
\begin{eqnarray}
E_{0+}^{\it (-)}(m_\pi) &=& \frac{m_N}{4\pi(m_N+m_\pi)}\frac{eG_A}{2F}
       \left[1+\frac{m_\pi^2}{8m_N^2}-\frac{m_\pi^2}{4m_N^2}(\mu_p-\mu_n)
       +\frac{\pi^2m_\pi^2}{4(4\pi{F})^2}
       \right.\nonumber \\
   &&  \left.-\frac{m_\pi^2}{G_A(4\pi{F})^2}\left(2b_{19}-2b_{21}^r(\mu)
       -2b_{22}^r(\mu)-b_{23}+G_A{\rm ln}\frac{m_\pi^2}{\mu^2}\right)
       \right] \; , \\
m_{1+}^{\it (-)}(m_\pi) &=& \frac{m_N}{4\pi(m_N+m_\pi)}\frac{eG_A}{2F}
       \left[-\frac{1}{6m_\pi^2}-\frac{1}{12m_\pi{m_N}}
       -\frac{(\mu_p-\mu_n)}{6m_\pi{m}_N}
       +\frac{5}{48m_N^2}\right.\nonumber \\
   &&  \left. -\frac{(\mu_p-\mu_n)}{6m_N^2}+\frac{2G_A^2}{3(4\pi{F})^2}
       -\frac{2G_A^2\pi}{3(4\pi{F})^2}+\frac{G_A^2\pi^2}{12(4\pi{F})^2}
       \right.\nonumber \\
   &&  \left. +\frac{1}{6G_A(4\pi{F})^2}
       \left(2b_{19}-4b_{22}^r(\mu)-2b_{23}-2G_A^3{\rm ln}\frac{m_\pi^2}{\mu^2}
       \right)\right] \; , \\
m_{1-}^{\it (-)}(m_\pi) &=& \frac{m_N}{4\pi(m_N+m_\pi)}\frac{eG_A}{2F}
       \left[\frac{1}{3m_\pi^2}+\frac{1}{6m_\pi{m_N}}
       -\frac{(\mu_p-\mu_n)}{6m_\pi{m}_N}
       +\frac{1}{24m_N^2} \right.\nonumber \\
   &&  \left. -\frac{(\mu_p-\mu_n)}{6m_N^2}
       -\frac{4G_A^2}{3(4\pi{F})^2} 
       -\frac{2G_A^2\pi}{3(4\pi{F})^2}
       +\frac{G_A^2\pi^2}{3(4\pi{F})^2} \right.\nonumber \\
   &&  \left. -\frac{1}{3G_A(4\pi{F})^2}
       \left(2b_{19}-4b_{22}^r(\mu)-2b_{23}-2G_A^3{\rm ln}\frac{m_\pi^2}{\mu^2}
       \right)\right] \; , \\
e_{1+}^{\it (-)}(m_\pi) &=& \frac{m_N}{4\pi(m_N+m_\pi)}\frac{eG_A}{2F}
       \left[\frac{1}{6m_\pi^2}+\frac{1}{12m_\pi{m_N}}-\frac{5}{48m_N^2}
       -\frac{b_{19}}{3G_A(4\pi{F})^2}\right] \; .\label{thresh8}
\end{eqnarray}

To make contact with previous work, observe that the $O(p)$ and $O(p^2)$ parts
of these expressions are just what one would obtain from an expansion of the
usual Born graphs using pseudovector coupling. The $O(p^3)$ parts contain
higher order pieces of the expansion of the Born graphs, loop contributions,
and contributions from the part of the Lagrangian involving the LEC's.

The numerical values of the threshold multipoles at each order in HBChPT are
displayed in Table~\ref{tab:thresh}.  The $O(p^3)$ results are given for both
solutions, A($n$) and B($n$). Again the results are essentially the same within
errors for any of the subsets of data used, though the fit is most accurate
when the full 35 points are included.   The $m_{1+}$ and $m_{1-}$ multipoles
differ dramatically between A($n$) and B($n$), as they have an important
dependence on $b_{10}$ which is quite different for the two solutions. $e_{1+}$
is constant, as it depends only on the parameter $b_{19}$ which was fixed from
muon capture \cite{FLMS} and $E_{0+}$ is nearly constant as it depends only on
the parameters $b_{22}^r, b_{21}^r, b_{19}$ and $b_{23}$ which are all
essentially the same for the two fits.

Also shown in Table~\ref{tab:thresh} are the results of a dispersion theory
calculation by Hanstein, Drechsel and Tiator.\cite{Tiator} For the electric
multipoles $E_{0+}$ and $e_{1+}$ the agreement with the HBChPT results is quite
good for both the $\pi^+$ and $\pi^-$ cases. For the magnetic multipoles
$m_{1+}$ and $m_{1-}$ the agreement with A(35) is good, albeit not
spectacular. One must recognize however that there are uncertainties in the
dispersion relations results also, which were quoted only for the $E_{0+}$
multipole. For the B(35) fit however the HBChPT and dispersion results for
these multipoles are quite different. Thus comparison with the dispersion
relation results strongly favors the A(35) solution over the B(35) one.

One can gain some further insight via a more detailed comparison with the
dispersion relation results. Observe first that
Eqs. (\ref{thresh1}-\ref{thresh8}) give the eight observable multipole
amplitudes in terms of four parameters: $b_{10}$, $b_{19}$, $b_{21}^r(\mu)$ and
$2b_{22}^r(\mu)+b_{23}$.  This means that four {\it parameter-free\/} relations
exist among the multipoles in the $O(p^3)$ HBChPT calculation.  For example,
Table~\ref{tab:nobs} gives a set of four quantities which are independent of
these four parameters, along with their values as obtained from HBChPT and
dispersion theory. For these four quantities the convergence of the HBChPT
expansion is good and the results agree quite well with the dispersion relation
predictions of Ref. \cite{Tiator}.

This idea can be carried a step further by looking at combinations of the
multipoles which depend on only one or only a few of the $b_i$'s. Such results
are tabulated in Table~\ref{tab:fewbs}. The multipole $e_{1+}^{\it (-)}$
depends, in fact only weakly, on $b_{19}$ and one can see from the table that
the HBChPT results converge well and agree with the dispersion theory result.
The next two entries, $m_{1+}^{\it (-)}$ and $m_{1-}^{\it (-)}$ depend in
addition on the combination $2 b_{22}^r + b_{23}$ and also agree with the
dispersion relation results.  The next entry $E_{0+}^{\it (-)}$ depends in
addition on $b_{21}^r$ and the following one, $E_{0+}^{\it (-)}+3 m_\pi^2
(m_{1+}^{\it (-)}-e_{1+}^{\it (-)})$ depends only on $b_{21}^r$. Both show good
convergence and good agreement with the dispersion theory. Finally the last two
entries $m_{1+}^{\it (0)}$ and $m_{1-}^{\it (0)}$ depend only on $b_{10}$. Here
the B(35) solution is clearly ruled out by comparison with the dispersion
relation results. The A(35) solution agrees moderately well, especially since
the dispersion results come from taking the difference of two large numbers,
and so probably have significant uncertainties.  As found before however, the
convergence of the magnetic multipoles is not as good as for the electric
multipoles.
   
One can summarize the results of this evaluation of the threshold multipoles
and comparison with the dispersion relation calculation of Ref. \cite{Tiator}
as follows. Generally the HBChPT calculation produces results for the
multipoles for the physical processes that converge and that agree with the
dispersion relation calculation. Likewise the various LEC's seem to be well
determined. The second solution, B(35), which could not be distinguished from
the other one on the basic of $\chi^2$ alone, seems to be ruled out by
comparison with dispersion relation results. The weakest link appears to be in
the convergence of the HBChPT expansion for the magnetic multipoles, which is
not as good as that for the electric multipoles, and in the detailed
combinations of multipoles depending on $b_{10}$ alone.

To improve the calculation it might be interesting to extend it to one higher
order, which can be done still within the context of a one-loop
calculation. Thus one could see if the $O(p^4)$ terms indicated real
convergence. One might also think about including the $\Delta(1232)$ as an
explicit degree of freedom. In the present calculation $\Delta$ effects are
included implicitly in the LEC's, which is a perfectly consistent approach. One
alternatively could extract them explicitly along the lines of
Ref. \cite{Thomas}. Very preliminary estimates seem to indicate that such
effects are relatively small in the very near threshold region we are
considering, but it might be worth doing a full calculation.

Finally, as somewhat of a side issue, we note that an alternative
representation of the near-threshold differential cross section which is often
used is
\begin{eqnarray}
   \frac{\omega}{|\vec{q}|}\frac{{\rm d}\sigma^{\gamma N \rightarrow \pi N}}
   {{\rm d}\Omega_\pi} &=& A+Bx+Cx^2 \\
   A &=& |E_{0+}|^2+\frac{1}{2}|P_2|^2+\frac{1}{2}|P_3|^2 \\
   B &=& 2{\rm Re}(E_{0+}P_1^*) \\
   C &=& |P_1^2|-\frac{1}{2}|P_2|^2-\frac{1}{2}|P_3|^2 \\
   P_1 &=& 3E_{1+}+M_{1+}-M_{1-} \\
   P_2 &=& 3E_{1+}-M_{1+}+M_{1-} \\
   P_3 &=& 2M_{1+}+M_{1-}
\end{eqnarray}
However, this near-threshold result differs somewhat from the general result we
have used. It is obtained by expanding the original amplitude, e.g. the pion
pole contributions, and keeping terms only through $x^2$, which is sufficient
to give the cross section in terms of s- and p-wave multipoles. In contrast we
used the square of the full HBChPT amplitude to get the cross section, and only
later after fitting the data extracted the s- and p-wave multipoles.

\section{Summary and Outlook}\label{sec5}

We have investigated the radiative capture of a charged pion by a nucleon using
heavy baryon chiral perturbation theory and have obtained explicit expressions
for the amplitude and for the s- and p-wave multipoles, expressed, as is more
conventional, as amplitudes for the inverse photoproduction process.  Up to
$O(p^3)$, these expressions depend upon three parameters that were determined
by fitting to data for $\pi^-$ capture by a proton and for very near threshold
photoproduction.  Two satisfactory fits were obtained, which were
indistinguishable, based only on comparison with the data.

Using the LEC's obtained from these fits, the eight s- and p-wave multipoles
(four for the $\pi^+$ case and four for the $\pi^-$ case) were calculated and
compared with results previously obtained from dispersion theory
\cite{Tiator}. In general the agreement was good for one of the fits, A(35),
whereas there were significant differences when the other fit was used. This
same result held for combinations of the multipoles depending on just a few of
the parameters.  We thus conclude that the A(35) fit gives an acceptable
result, and thus that the three parameters determined in that fit, $b_{10},
b_{21}^r, b_{22}^r$ and given in Table \ref{tab:bs} are available for future
studies of other reactions.

In general the convergence of the HBChPT expansion was very good for the
electric multipoles, but somewhat less good for the magnetic ones.  This
suggests that it might be valuable to consider extending the present work to
$O(p^4)$ or to include explicit $\Delta(1232)$ fields in the chiral Lagrangian.

\section*{Acknowledgments}

We are grateful to Dave Hutcheon for providing us with unpublished TRIUMF data,
to Elie Korkmaz for the data from the SAL experiment, and to Dave Hutcheon,
Lothar Tiator, and Olaf Hanstein for helpful conversations.  This work was
supported in part by the Natural Sciences and Engineering Research Council of
Canada.

\bigskip
\begin{appendix}
\section{Structure amplitudes}\label{appA}
Up to $O(p^3)$ in HBChPT, the structure amplitudes of Eq. (\ref{eq:T}),
corresponding to the photoproduction process $\gamma + N \rightarrow \pi + N$,
are found to be
\begin{eqnarray}
F_1^{\it (0)}(E_\pi,x) &=& \frac{m_N}{4\pi\sqrt{s}}\frac{eG_A}{2F}
       \left\{\frac{1}{2m_N}\left[-E_\pi+x|\vec{q}|(\mu_p+\mu_n)\right]
       +\frac{2xE_\pi|\vec{q}|b_{10}}{G_A(4{\pi}F)^2}
       \right.\nonumber \\
  &&   \left.+\frac{1}{4m_N^2}\left[-|\vec{q}|^2 - \frac{1}{2}xE_\pi|\vec{q}|
       + (2E_\pi^2-m_\pi^2+xE_\pi|\vec{q}|-2x^2|\vec{q}|^2)(\mu_p+\mu_n)\right]
       \right\} \; , \\
F_1^{\it (-)}(E_\pi,x) &=& \frac{m_N}{4\pi\sqrt{s}}\frac{eG_A}{2F}
	\left\{1-\frac{x|\vec{q}|}{2m_N}(\mu_p-\mu_n) \right.\nonumber \\
  &&   \left. +\frac{1}{4m_N^2}\left[+E_\pi^2-\frac{m_\pi^2}{2}
       -(2E_\pi^2-m_\pi^2+xE_\pi|\vec{q}|-2x^2|\vec{q}|^2)(\mu_p-\mu_n)
      \right] \right. \nonumber \\
  &&   \left. -\frac{2m_\pi^2b_{19}}{G_A(4{\pi}F)^2}
       +\frac{2E_\pi^2}{G_A(4{\pi}F)^2}\left(b_{21}^r(\mu)-\frac{G_A}{2}
       (1+G_A^2){\rm ln}\frac{m_\pi^2}{\mu^2}\right) \right. \nonumber \\
  &&   \left. +\frac{E_\pi(E_\pi-x|\vec{q}|)}{G_A(4{\pi}F)^2}
       \left(2b_{22}^r(\mu)+b_{23}+G_A^3{\rm ln}\frac{m_\pi^2}{\mu^2}\right)
       +\frac{1}{4(4{\pi}F)^2}\left[\pi^2m_\pi^2 \rule[-2ex]{0cm}{4ex}
       \right. \right. \nonumber \\
   &&  \left. \left.  -8E_\pi|\vec{q}|{\rm ln}\left(\frac{E_\pi+|\vec{q}|}
       {m_\pi}\right)
       +4i{\pi}m_\pi^2{\rm ln}\left(\frac{E_\pi+|\vec{q}|}{m_\pi}\right)
       -4m_\pi^2\left({\rm ln}\left(\frac{E_\pi+|\vec{q}|}{m_\pi}\right)
       \right)^2 \right. \right.  \nonumber \\
   &&  \left. \left. +4i\pi{E}_\pi|\vec{q}| \rule[-2ex]{0cm}{4ex}
       \right]
       +\frac{xG_A^2E_\pi|\vec{q}|}{(4{\pi}F)^2}\left[2
       -\frac{2|\vec{q}|}
	{E_\pi}{\rm ln}\left(\frac{E_\pi+|\vec{q}|}{m_\pi}\right)
       +\frac{\pi^2m_\pi^2}{4E_\pi^2}
       -\frac{2\pi{m}_\pi}{E_\pi} \right.\right. \nonumber \\
   &&  \left.\left.+\frac{m_\pi^2}{E_\pi^2}
	\left({\rm ln}\left(\frac{E_\pi+|\vec{q}|}{m_\pi}\right)\right)^2
       \right] \right\} \; , \\
F_2^{\it (0)}(E_\pi,x) &=& \frac{m_N}{4\pi\sqrt{s}}\frac{eG_A}{2F}
       \left\{\frac{E_\pi |\vec{q}|}{8 m_N^2}
       -\frac{x |\vec{q}|^2}{4 m_N^2}(\mu_p+\mu_n)
       +\frac{2E_\pi|\vec{q}|b_{10}}{G_A(4{\pi}F)^2}\right\} \; , \\
F_2^{\it (-)}(E_\pi,x) &=& \frac{m_N}{4\pi\sqrt{s}}\frac{eG_A}{2F}
       \left\{-\frac{|\vec{q}|}{2m_N}(\mu_p-\mu_n)
       +\frac{|\vec{q}|}{4m_N^2}\left[E_\pi-(E_\pi-x|\vec{q}|)(\mu_p-\mu_n)
       \right] \right.\nonumber \\
  &&   \left. +\frac{G_A^2E_\pi|\vec{q}|}{2(4{\pi}F)^2}
       \left[\frac{\pi^2m_\pi^2}{E_\pi^2}-\frac{4\pi{m}_\pi}{E_\pi}
       -2{\pi}i\frac{|\vec{q}|}{E_\pi}
       +2{\pi}i\frac{m_\pi^2}{E_\pi^2}{\rm ln}
       \left(\frac{E_\pi+|\vec{q}|}{m_\pi}
       \right)\right]\right\} \; , \\
F_3^{\it (0)}(E_\pi,x) &=& \frac{m_N}{4\pi\sqrt{s}}\frac{eG_A}{2F}
       \left\{-\frac{|\vec{q}|}{2m_N}(\mu_p+\mu_n)
       +\frac{E_\pi|\vec{q}|}{8m_N^2}\left[3-2(\mu_p+\mu_n)\right]
       \right. \nonumber \\
   &&  \left.  +\frac{x|\vec{q}|^2}{2m_N^2}(\mu_p+\mu_n) 
       -\frac{2E_\pi|\vec{q}|b_{10}}{G_A(4{\pi}F)^2} \right\} \; , \\
F_3^{\it (-)}(E_\pi,x) &=& \frac{m_N}{4\pi\sqrt{s}}\frac{eG_A}{2F}
       \left\{+\frac{|\vec{q}|}{(E_\pi-x|\vec{q}|)}
       +\frac{|\vec{q}|}{2m_N}(\mu_p-\mu_n)
       -\frac{m_\pi^2|\vec{q}|}{4m_N^2(E_\pi-x|\vec{q}|)}
       \right.\nonumber \\
   &&  \left. +\frac{|\vec{q}|}{m_N^2}\left[-\frac{E_\pi}{4}
       +\frac{m_\pi^2}{8(E_\pi-x|\vec{q}|)}
       +\frac{1}{4}(E_\pi-2x|\vec{q}|)(\mu_p-\mu_n)\right] 
       -\frac{2G_A^2E_\pi|\vec{q}|}{(4{\pi}F)^2}
       \right. \nonumber \\
   &&  \left. -\frac{2m_\pi^2|\vec{q}|b_{19}}
	{G_A(4{\pi}F)^2(E_\pi-x|\vec{q}|)} 
       +\frac{E_\pi|\vec{q}|}{G_A(4{\pi}F)^2}
       \left(2b_{22}^r(\mu)+b_{23}+G_A^3{\rm ln}\frac{m_\pi^2}{\mu^2}\right)
       \right. \nonumber \\
   &&  \left. +\frac{G_A^2E_\pi|\vec{q}|}{(4{\pi}F)^2}\left[\frac{2|\vec{q}|}
	{E_\pi}{\rm ln}\left(\frac{E_\pi+|\vec{q}|}{m_\pi}\right)
       -\frac{\pi^2m_\pi^2}{4E_\pi^2}
       -\frac{m_\pi^2}{E_\pi^2}
	\left({\rm ln}\left(\frac{E_\pi+|\vec{q}|}{m_\pi}\right)\right)^2
        +\frac{2\pi{m}_\pi}{E_\pi}
       \right] \right\} \; , \\
F_4^{\it (0)}(E_\pi,x) &=& \frac{m_N}{4\pi\sqrt{s}}\frac{eG_A}{2F}
       \left\{\frac{|\vec{q}|^2}{2m_NE_\pi}
       +\frac{1}{m_N^2}\left[\frac{E_\pi^2}{4}-\frac{m_\pi^4}{4E_\pi^2}
       -\frac{x|\vec{q}|^3}{2E_\pi}-\frac{|\vec{q}|^2}{4}(\mu_p+\mu_n)
       \right]\right\} \; , \\
F_4^{\it (-)}(E_\pi,x) &=& \frac{m_N}{4\pi\sqrt{s}}\frac{eG_A}{2F}
       \left\{-\frac{|\vec{q}|^2}{E_\pi(E_\pi-x|\vec{q}|)}
       -\frac{|\vec{q}|^2}{2m_NE_\pi}\left[1+\frac{m_\pi^2}
       {E_\pi(E_\pi-x|\vec{q}|)}\right] \right.\nonumber \\
   &&  \left. -\frac{|\vec{q}|^2}{4m_N^2}\left[1+\frac{m_\pi^2}{E_\pi^2}
       -\frac{2x|\vec{q}|}{E_\pi}-(\mu_p-\mu_n)
       -\frac{3m_\pi^2}{2E_\pi(E_\pi-x|\vec{q}|)}
       +\frac{m_\pi^4}{E_\pi^3(E_\pi-x|\vec{q}|)} \right] \right. \nonumber \\
   &&  \left. +\frac{2m_\pi^2|\vec{q}|^2b_{19}}
       {G_A(4{\pi}F)^2E_\pi(E_\pi-x|\vec{q}|)} \right\} \; ,
\end{eqnarray}
where $|\vec{q}| = \sqrt{E_\pi^2-m_\pi^2}$,
$m_N$ is the renormalized nucleon mass, and $m_\pi$ is the renormalized
pion mass.

Note that all of the parameters in these expressions have been
renormalized.  The calculation was performed using the bare Lagrangian
parameters, which were then converted to renormalized parameters as follows:
\begin{eqnarray}
2a_7 &=& \mu_p + \mu_n \; , \\
4a_6 &=& \mu_p - \mu_n + \frac{4{\pi}G_A^2m_\pi{m}_N}{(4{\pi}F)^2} \; , \\
F_0 &=& F\left\{1-\frac{m_\pi^2}{F^2}\left[l_4^r(\mu)-
         \frac{1}{(4\pi)^2}{\rm ln}
	\left(\frac{m_\pi^2}{\mu^2}\right)\right]\right\} \; , \\
g_A &=& G_A - \frac{4a_3G_Am_\pi^2}{m_N^2} + \frac{G_A^3
	m_\pi^2}{(4{\pi}F)^2} - \frac{4m_\pi^2}{(4{\pi}F)^2}\left[b_{17}^r(\mu)
	-\frac{G_A}{4}(1+2G_A^2){\rm ln}\left(\frac{m_\pi^2}{\mu^2}\right)
	\right] \; .
\end{eqnarray}
$\mu_p \simeq 2.79$ and $\mu_n \simeq -1.91$ are the magnetic moments of the
proton and neutron, respectively.  The expression for the bare pion decay
constant $F_0$ in terms of the renormalized $F$ and for the bare $g_A$ in terms
of the physical $G_A \simeq 1.26$ depend somewhat on the explicit form of the
Lagrangian used, and are derived, for example, in Ref. \cite{FLMS}.

\end{appendix}

\bigskip

\newpage

\begin{table*}
   \caption{Values of the three coefficients in the $O(p^3)$ HBChPT Lagrangian
            which are obtained from a least-squares, three parameter, fit to
            various subsets of the experimental data.  In each case there are
            two roughly equivalent well defined minima of $\chi^2$ labeled by A
            and B. The data consists of (a) 11 measurements of $\pi^- p
            \rightarrow \gamma n$ with $T_\pi \leq 19.85$ MeV
            \protect\cite{newdata}; (b) 16 measurements of $\pi^- p \rightarrow
            \gamma n$ with $T_\pi \geq 27.4$ MeV \protect\cite{Salomon}; and
            (c) 8 measurements of $\gamma p \rightarrow n \pi^+$ at $T_\gamma
            \simeq 153$ MeV \protect\cite{Korkmaz}. The arguments of A and B
            correspond to the number of data in the set chosen so that 11
            $\sim$ set (a), 16 $\sim$ set (b), 27 $\sim$ sets (a)+(b), and 35
            $\sim$ sets (a)+(b)+(c).  As input, we use $b_{19} = -0.7 \pm 0.4$
            and $b_{23} = -3.1 \pm 0.3$ as determined in
            Ref. \protect\cite{FLMS}. Note that $b_{22}^r$ appears only in the
            combination $2 b_{22}^r +b_{23}$, so that its value obtained from
            fitting this data depends on the value taken for $b_{23}$. As
            discussed in the text, A(35) is considered to be the best result.
            }\label{tab:bs}
    \begin{tabular}{ccccc}
    & $\chi^2/{\rm d.o.f.}$ & $b_{10}$ & $b_{21}^r({\rm m_N})$ &
    $b_{22}^r({\rm m_N})$ \\
    \hline
    A(11) & 2.79 &   8.8$\pm$16.1 & -8.2$\pm$1.1 & 9.2$\pm$1.1 \\
    A(16) & 1.12 &   6.1$\pm$ 9.1 & -7.6$\pm$1.1 & 9.3$\pm$0.8 \\
    A(27) & 1.62 &  11.9$\pm$ 5.4 & -8.2$\pm$0.7 & 9.3$\pm$0.6 \\
    A(35) & 1.59 &  13.7$\pm$ 4.5 & -8.2$\pm$0.7 & 9.2$\pm$0.6 \\
    B(11) & 2.81 & -40.5$\pm$15.7 & -8.2$\pm$1.1 & 9.3$\pm$1.0 \\
    B(16) & 1.15 & -36.2$\pm$ 9.2 & -7.6$\pm$1.0 & 9.4$\pm$0.8 \\
    B(27) & 1.63 & -42.4$\pm$ 5.3 & -8.3$\pm$0.7 & 9.4$\pm$0.7 \\
    B(35) & 1.67 & -45.6$\pm$ 4.6 & -8.4$\pm$0.7 & 9.4$\pm$0.7
    \end{tabular}
\end{table*}
\begin{table*}
    \caption{Threshold s- and p-wave multipoles for the reactions
	     $\gamma n \rightarrow \pi^-p$ and $\gamma p \rightarrow \pi^+n$.
	     The $E_{0+}$ are in units of $10^{-3}/m_{\pi^+}$ and
	     the reduced $p$-wave multipoles are in units of
	     $10^{-3}/m_{\pi^+}^3$.  Dispersion theory results are
	     quoted from Ref.~\protect\cite{Tiator}.}\label{tab:thresh}
    \begin{tabular}{llllcccc}
    & & & & $E_{0+}$ & $m_{1+}$ & $m_{1-}$ & $e_{1+}$ \\
    \hline
    $\gamma n \rightarrow \pi^-p$
    & HBChPT & $O(p)$	&	& -28.2 & 4.7 & -9.4 & -4.7 \\
    &	     & $O(p^2)$ &	& -30.3 & 9.0 & -8.1 & -5.1 \\
    &	     & $O(p^3)$ & A(11) & -32.2$\pm$1.0 & 12.6$\pm$3.5
		  & -7.4$\pm$3.5 & -5.04$\pm$0.04 \\
    &	     & $O(p^3)$ & A(16) & -32.7$\pm$0.9 & 12.1$\pm$2.0
		  & -8.0$\pm$2.0 & -5.04$\pm$0.04 \\
    &	     & $O(p^3)$ & A(27) & -32.3$\pm$0.7 & 13.3$\pm$1.2
		  & -6.8$\pm$1.2 & -5.04$\pm$0.04 \\
    &	     & $O(p^3)$ & A(35) & -32.2$\pm$0.7 & 13.7$\pm$1.0
		  & -6.4$\pm$1.0 & -5.04$\pm$0.04 \\
    &	     & $O(p^3)$ & B(11) & -32.3$\pm$1.0 & 2.0$\pm$3.4
		  & -18.1$\pm$3.4 & -5.04$\pm$0.04 \\
    &	     & $O(p^3)$ & B(16) & -32.7$\pm$0.9 & 2.9$\pm$2.0
		  & -17.2$\pm$2.0 & -5.04$\pm$0.04 \\
    &	     & $O(p^3)$ & B(27) & -32.3$\pm$0.7 & 1.6$\pm$1.2
		  & -18.6$\pm$1.2 & -5.04$\pm$0.04 \\
    &	     & $O(p^3)$ & B(35) & -32.2$\pm$0.7 & 0.9$\pm$1.0
		  & -19.2$\pm$1.0 & -5.04$\pm$0.04 \\
    & \multicolumn{3}{l}{Dispersion theory}
		  & -31.7$\pm$0.2 & 11.2 & -8.3 & -4.9 \\
    \hline
    $\gamma p \rightarrow \pi^+n $
    & HBChPT & $O(p)$	&	& 28.2 & -4.7 & 9.4 & 4.7 \\
    &	     & $O(p^2)$ &	& 26.1 & -7.7 & 5.6 & 5.1 \\
    &	     & $O(p^3)$ & A(11) & 28.3$\pm$1.0 & -7.5$\pm$3.5
		  & 8.8$\pm$3.5 & 5.09$\pm$0.04 \\
    &	     & $O(p^3)$ & A(16) & 28.8$\pm$0.9 & -8.1$\pm$2.0
		  & 8.2$\pm$2.0 & 5.09$\pm$0.04 \\
    &	     & $O(p^3)$ & A(27) & 28.4$\pm$0.7 & -6.9$\pm$1.2
		  & 9.5$\pm$1.2 & 5.09$\pm$0.04 \\
    &	     & $O(p^3)$ & A(35) & 28.3$\pm$0.7 & -6.5$\pm$1.0
		  & 9.8$\pm$1.0 & 5.09$\pm$0.04 \\
    &	     & $O(p^3)$ & B(11) & 28.3$\pm$1.0 & -18.2$\pm$3.4
		  & -1.9$\pm$3.4 & 5.09$\pm$0.04 \\
    &	     & $O(p^3)$ & B(16) & 28.8$\pm$0.9 & -17.3$\pm$2.0
		  & -0.9$\pm$2.0 & 5.09$\pm$0.04 \\
    &	     & $O(p^3)$ & B(27) & 28.4$\pm$0.7 & -18.6$\pm$1.2
		  & -2.2$\pm$1.2 & 5.09$\pm$0.04 \\
    &	     & $O(p^3)$ & B(35) & 28.2$\pm$0.7 & -19.3$\pm$1.0
		  & -2.9$\pm$1.0 & 5.09$\pm$0.04 \\
      & \multicolumn{3}{l}{Dispersion theory}
		  & 28.0$\pm$0.2 & -9.6 & 6.1 & 4.9
    \end{tabular}
\end{table*}
\begin{table*}
    \caption{Four combinations of s- and p-wave threshold multipoles that are
	     independent of all $b_i$ parameters (including $b_{19}$) up to and
	     including $O(p^3)$ in HBChPT.  Their values are compared to the
	     dispersion theory results of Ref.~\protect\cite{Tiator}.  The
	     $E_{0+}$ are in units of $10^{-3}/m_{\pi^+}$ and the reduced
	     $p$-wave multipoles are in units of $10^{-3}/m_{\pi^+}^3$.
    }\label{tab:nobs}
    \begin{tabular}{ccccc}
    & \multicolumn{3}{c}{HBChPT} & dispersion theory \\
    & $O(p)$ & $O(p^2)$ & $O(p^3)$ & \\
    \hline
    $E_{0+}^{\it (0)}$ & 0 & -1.5 & -1.4	 & -1.3$\pm$0.1 \\
    $e_{1+}^{\it (0)}$ & 0 &  0	 &  0.02 &  0		\\
    $m_{1+}^{\it (0)} - m_{1-}^{\it (0)}$
			      & 0 &  1.3 &  1.3	 &  1.3 \\
    $2m_{1+}^{\it (-)} + m_{1-}^{\it (-)}$
			      & 0 & -7.0 & -8.5	 &  -9.6
    \end{tabular}
\end{table*}

\begin{table*}
    \caption{Some additional combinations of s- and p-wave multipoles that
             depend on subsets of the $b_i$ parameters up to and including
             $O(p^3)$ in HBChPT.  Their values are compared to the dispersion
             theory results of Ref.~\protect\cite{Tiator}.  The $E_{0+}$ are in
             units of $10^{-3}/m_{\pi^+}$ and the reduced $p$-wave multipoles
             are in units of $10^{-3}/m_{\pi^+}^3$.
    }\label{tab:fewbs}
    \begin{tabular}{ccrrrcrc}
    & Depends on & \multicolumn{3}{c}{HBChPT} &&  \multicolumn{2}{c} 
      {Dispersion theory} \\
    & & $O(p)$ & $O(p^2)$ & \multicolumn{1}{c}{$O(p^3)$} && 
      \phantom{xxxxxxxxx}&  \\
    \hline
    $e_{1+}^{\it (-)}$ & $b_{19}$ & 3.3 & 3.6 & 3.58$\pm$0.03 && 3.5 \\
    $m_{1+}^{\it (-)}$ &$b_{19},2 b_{22}^r+b_{23}$&-3.3 &  -5.9&
       -7.1$\pm$0.1 & & -7.4  \\
    $m_{1-}^{\it (-)}$ &$b_{19},2 b_{22}^r+b_{23}$&6.7 & 4.8&
       5.7$\pm$0.2  & & 5.1  \\
    $E_{0+}^{\it (-)}$ &$b_{19},2 b_{22}^r+b_{23},b_{21}^r$&20.0 & 20.0&
       21.4$\pm$0.5  & & 21.1  \\
    $E_{0+}^{\it (-)}+3 m_\pi^2 (m_{1+}^{\it (-)}-e_{1+}^{\it (-)})$ &
       $b_{21}^r$&0 & -8.5& -10.7$\pm$0.3  & & -11.4  \\
    $m_{1+}^{\it (0)}$ &$b_{10}$
			      & 0 &  0.4 & 2.6$\pm$0.7	 &A(35)&  0.6 \\
    & & & & -6.5$\pm$0.7 &B(35)&  \\
    $m_{1-}^{\it (0)}$ &$b_{10}$
			      & 0 & -0.9 & 1.2$\pm$0.7	 &A(35)&  -0.8 \\
    & & & & -7.8$\pm$0.7 &B(35)& 
    \end{tabular}
\end{table*}

\begin{figure}[tbh]
\caption{The cross section for the pion capture reaction, quoted as the reduced
         center-of-mass cross section for the inverse $\gamma n \rightarrow
         \pi^- p$ or [for (f)] $\gamma p \rightarrow \pi^+ n$ reaction.
         Experimental data are compared to the HBChPT predictions at $O(p)$
         (dotted line), $O(p^2)$ (dashed line), and $O(p^3)$ (solid line). The
         $O(p^3)$ result corresponds to A(35) and B(35) which are
         indistinguishable in these plots. \\
	 (a)~$T_\pi = 9.88$~MeV,
	 (b)~$T_\pi = 14.62$~MeV,
	 (c)~$T_\pi = 19.85$~MeV,
	 (d)~$T_\pi = 27.4$~MeV,
	 (e)~$T_\pi = 39.3$~MeV,
         (f)~$T_\gamma = 153$~MeV, $T_\pi = 3.06$~MeV.
	 }\label{plots}
\end{figure}
\begin{figure}[tbh]
\epsfxsize=380pt \epsfbox[30 419 498 732]{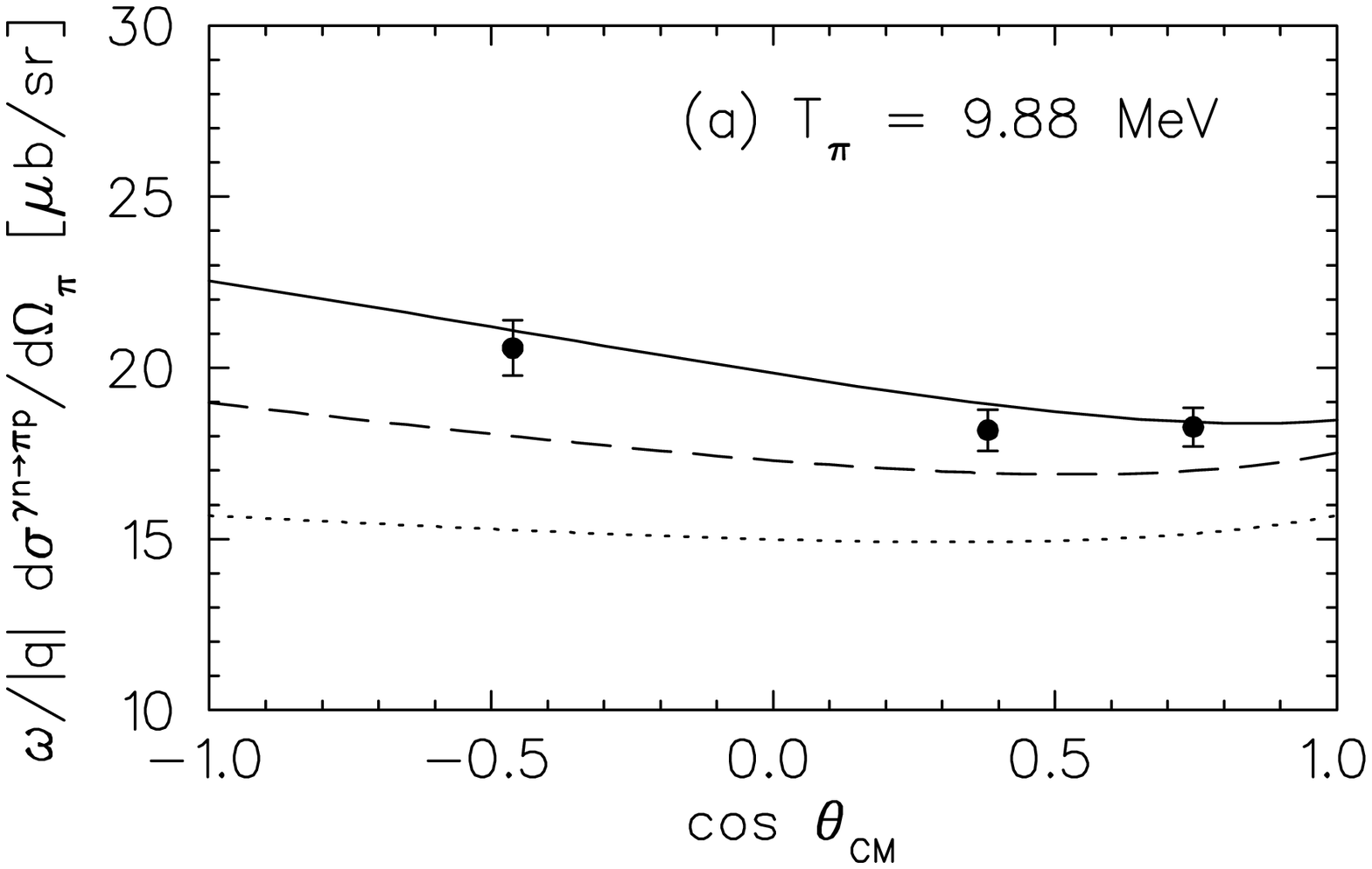}
\end{figure}
\begin{figure}[tbh]
\epsfxsize=380pt \epsfbox[30 419 498 732]{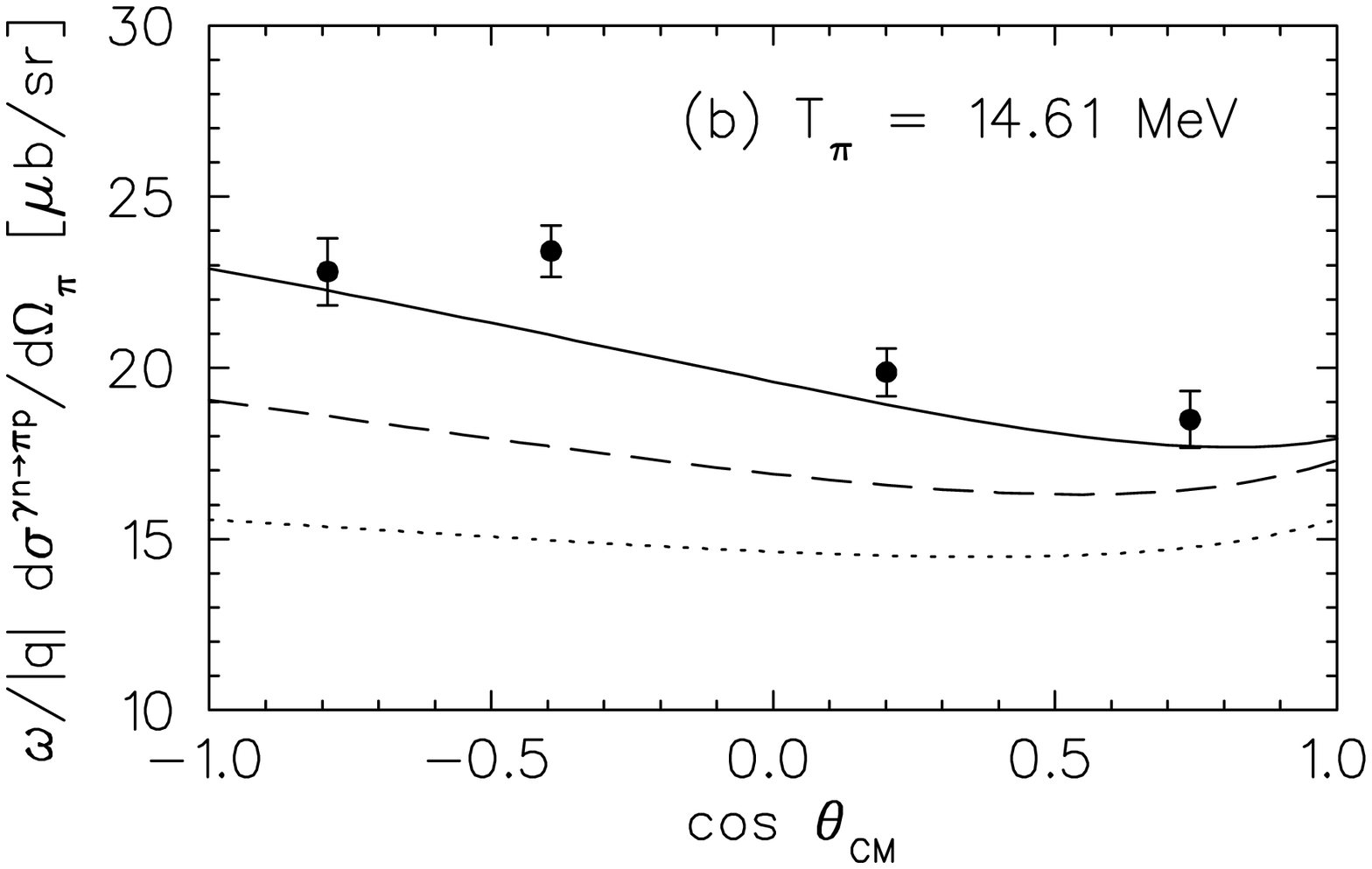}
\end{figure}
\begin{figure}[tbh]
\epsfxsize=380pt \epsfbox[30 419 498 732]{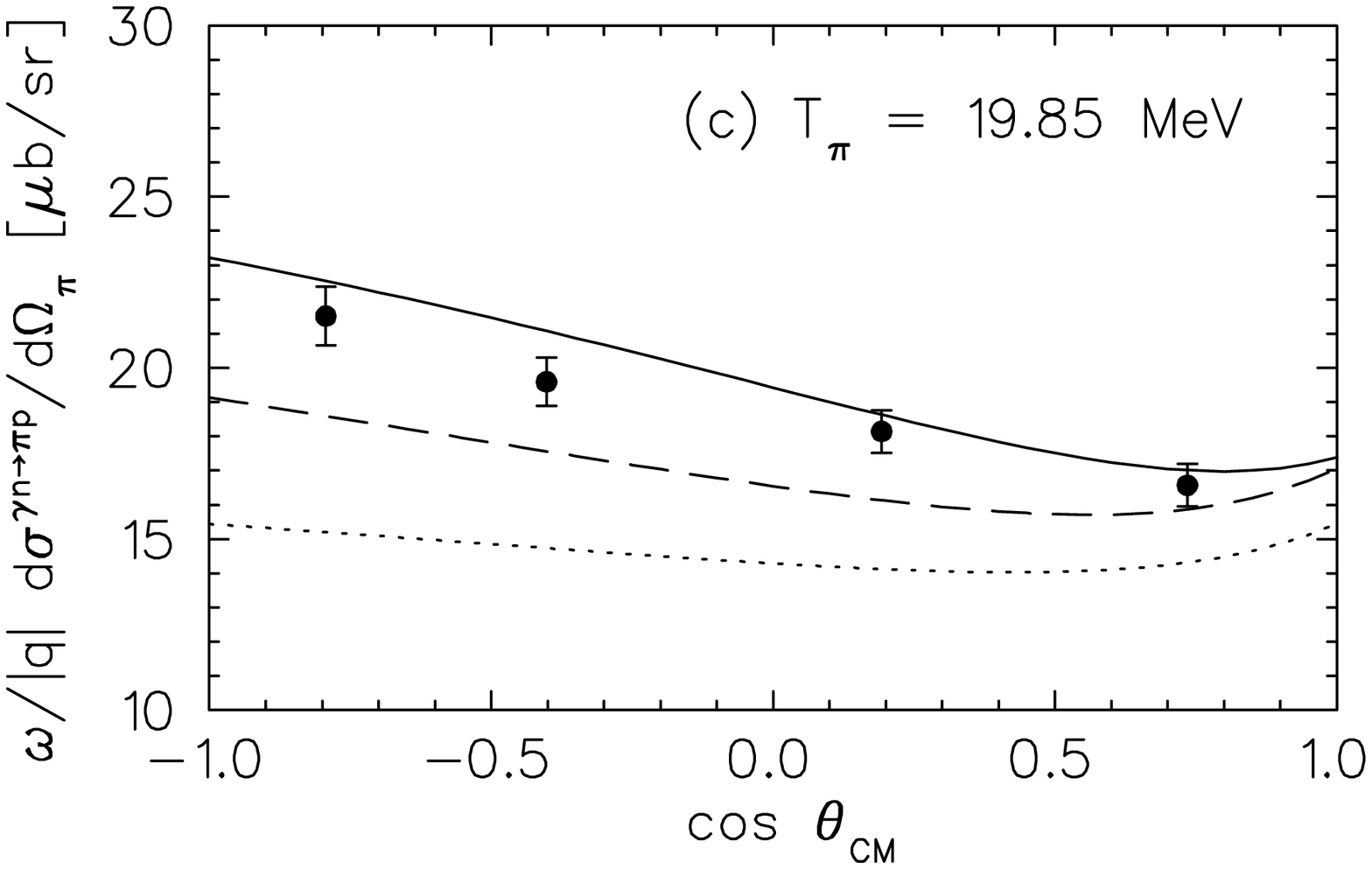}
\end{figure}
\begin{figure}[tbh]
\epsfxsize=380pt \epsfbox[30 419 498 732]{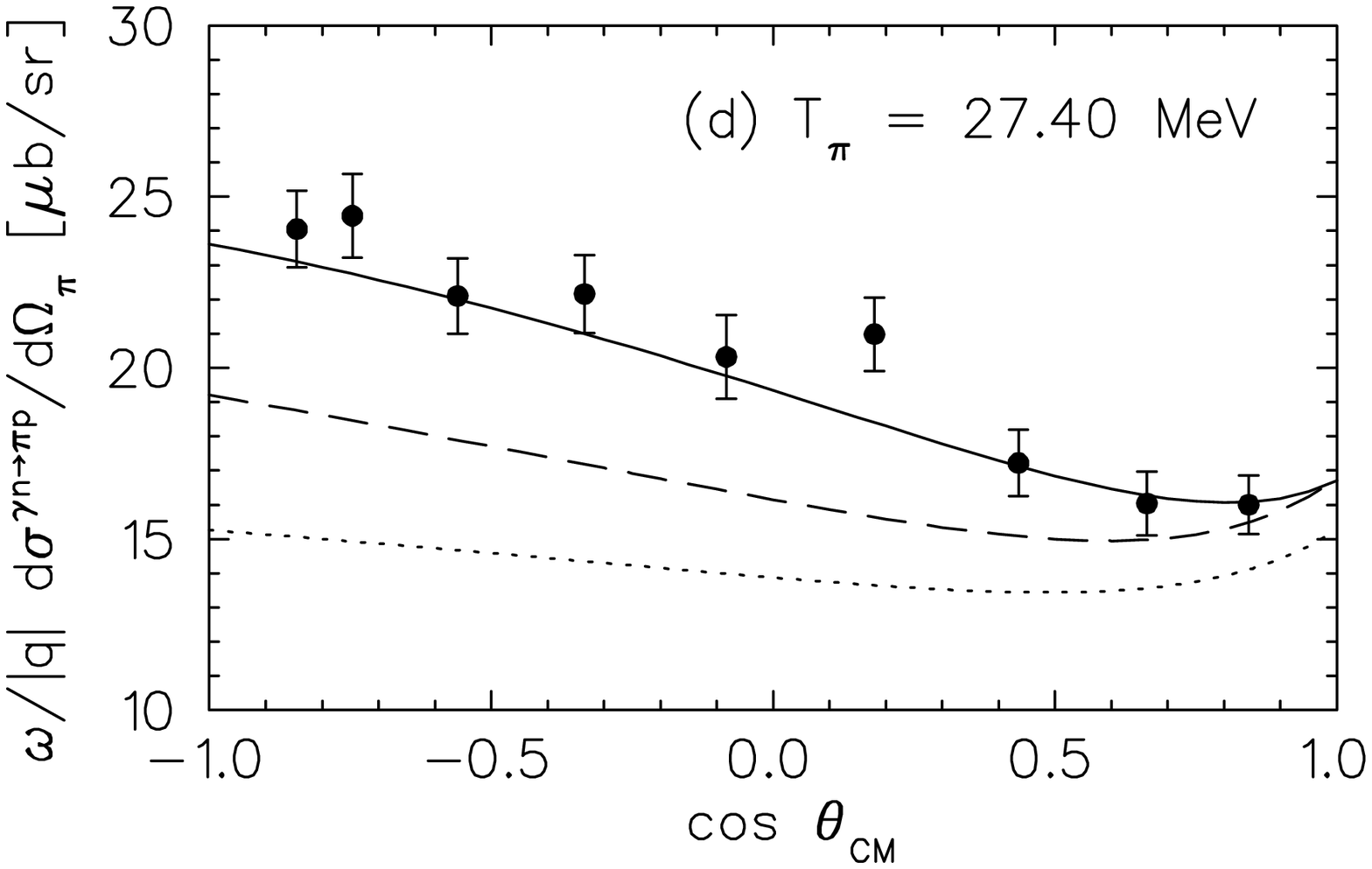}
\end{figure}
\begin{figure}[tbh]
\epsfxsize=380pt \epsfbox[30 419 498 732]{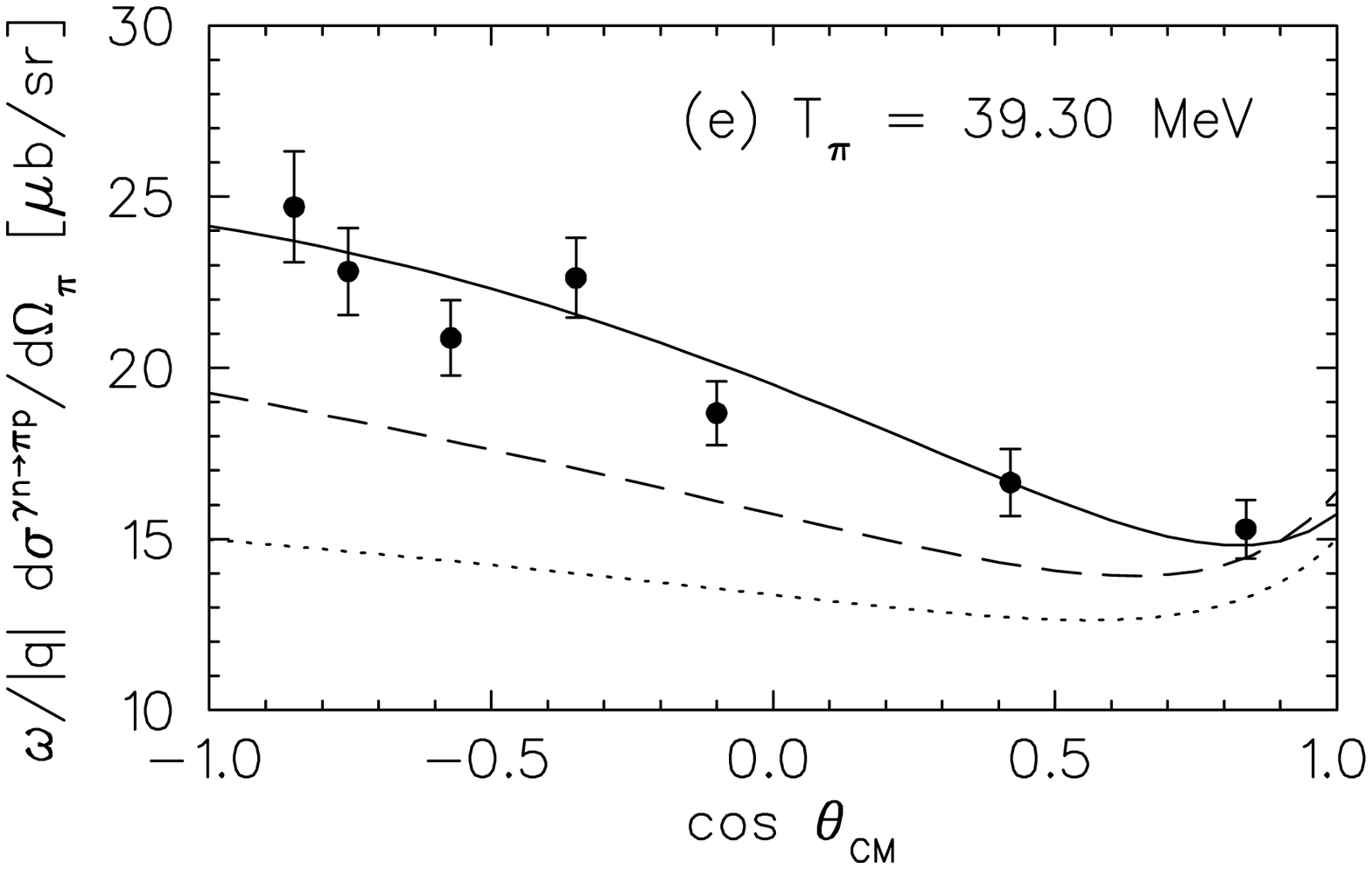}
\end{figure}
\begin{figure}[tbh]
\epsfxsize=380pt \epsfbox[30 419 498 732]{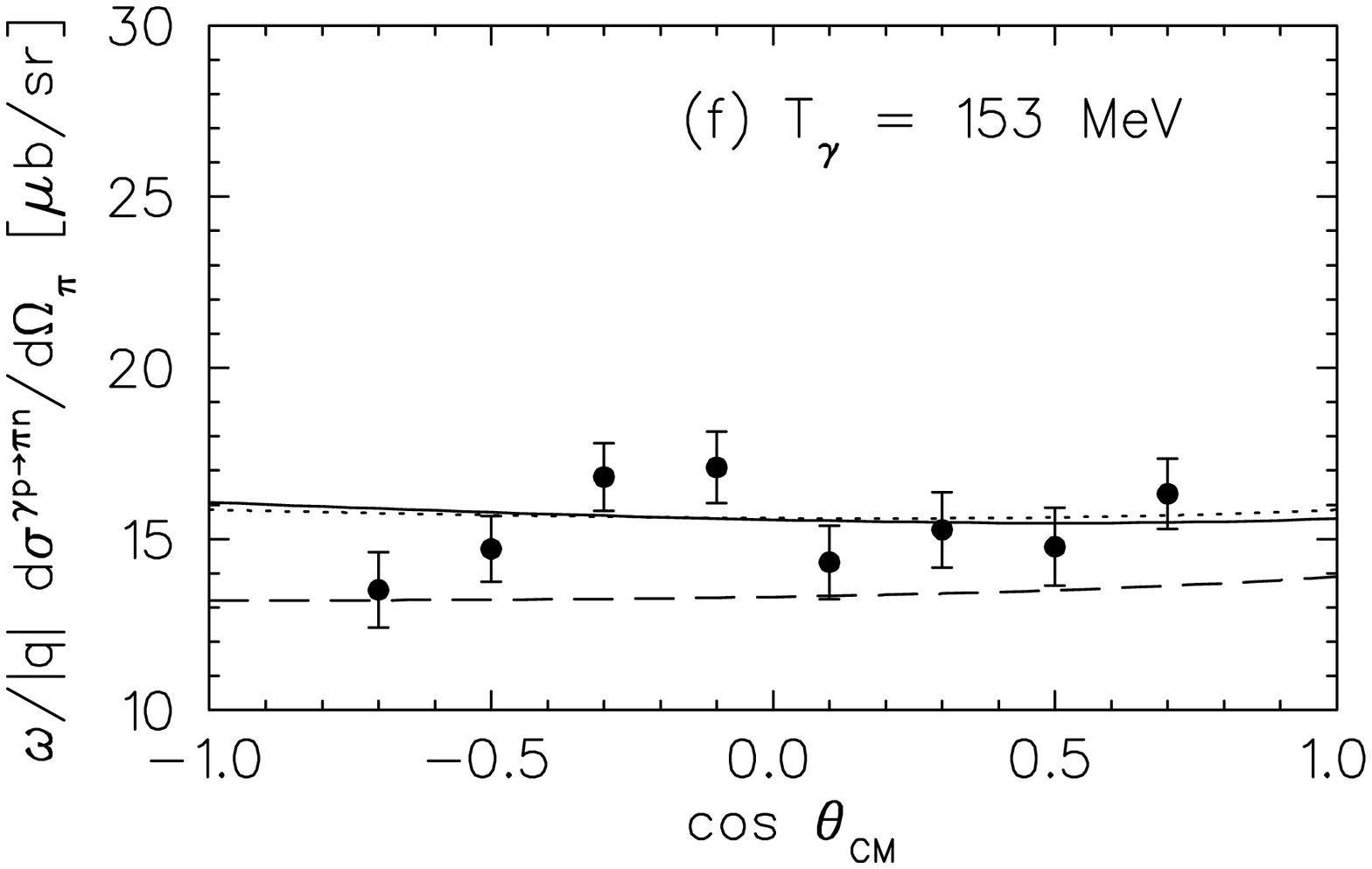}
\end{figure}

\end{document}